\documentclass[iop]{emulateapj}

\usepackage{amsmath}
\usepackage[usenames,dvipsnames]{color}
\usepackage{amssymb}
\usepackage{natbib}
\usepackage{bm}

\begin{document}

\title{Anomalous circular polarization profiles in the \ion{He}{1} 1083.0~nm multiplet from solar spicules}

%\author{M. J. Mart\'{\i}nez Gonz\'alez\altaffilmark{1}, A. Asensio Ramos, R. Manso Sainz, C. Beck, \& L. Belluzzi}
\author{M. J. Mart\'{\i}nez Gonz\'alez, A. Asensio Ramos, R. Manso Sainz, C. Beck, \& L. Belluzzi}
\affil{Instituto de Astrof\'{\i}sica de Canarias, 38205, La Laguna, Tenerife, Spain \\
Departamento de Astrof\'{\i}sica, Universidad de La Laguna, La Laguna, Tenerife, Spain \email{marian@iac.es}}

%\altaffiltext{1}{MJMG, CB, and RMS (PI) made the observations; MJMG and CB reduced the data; MJMG and AAR made the inversions; RMS, AAR, and LB worked out the atomic orientation scenarios; MJMG, AAR, and RMS wrote the paper. All authors discussed extensively the results and contributed to the final version of the paper.}

% % abstract cannot exceed 300 words
\begin{abstract}
We report Stokes vector observations of solar spicules and a prominence 
in the \ion{He}{1} 1083 nm multiplet carried out with
the Tenerife Infrared Polarimeter. The observations show linear polarization profiles that are produced by scattering processes in the presence of a magnetic field.
After a careful data reduction, we demonstrate the existence of extremely asymmetric Stokes $V$ profiles in the spicular material that we are able to model with two magnetic components along the line of sight, and under the presence of atomic orientation in the energy levels that give rise to the multiplet. We discuss some possible scenarios that can generate the atomic orientation in spicules. We stress the importance of spectropolarimetric observations across the limb to distinguish such signals from observational artifacts.
\end{abstract}

\maketitle

%%%%%%%%%%%%%%%%%%%%%%%%%%%%%%%%%%%%%%%%
%%%%%%%%%%%%%%%%%%%%%%%%%%%%%%%%%%%%%%%%
% INTRODUCTION
%%%%%%%%%%%%%%%%%%%%%%%%%%%%%%%%%%%%%%%%
%%%%%%%%%%%%%%%%%%%%%%%%%%%%%%%%%%%%%%%%

\section{Introduction}
Spectropolarimetry in the neutral helium multiplets at 1083.0~nm and 587.6~nm ($D_3$ line) is one of the most valuable tools (and one of the few available) for the quantitative investigation of the magnetic properties of plasma structures embedded in the chromosphere and corona, such as spicules and prominences \citep{landi_d3_82,querfeld85,bommier94,casini03,lopezariste_casini05,ramelli06_2,ramelli06, harvey_hall71,ruedi96,lin98,trujillo_nature02,Lagg04,trujillo_merenda05,socas_elmore05,merenda06,centeno+10}.
The linear polarization signals in both multiplets are generated by the joint action of the transversal Zeeman effect and scattering polarization modified by the Hanle effect \citep[e.g.,][]{trujillo_asensio07}.
However, the circular polarization signal of the 1083.0~nm multiplet is always dominated by the longitudinal Zeeman effect alone, producing characteristic antisymmetric profiles, whereas in $D_3$ it has an important contribution from the alignment-to-orientation conversion mechanism \citep[A-O mechanism;][]{kemp84,landi_landolfi04}, which generates a significant amount of net circular polarization (NCP) in the line \citep{landi_d3_82}. 
The A-O mechanism operates when the Larmor frequency corresponding to the magnetic field is of the order of the energy separation (in frequency units) between the fine structure levels of a given spectral term. 
In the D$_3$ line this occurs for magnetic fields stronger than 10~G; in the 1083.0~nm line, it is negligible for fields below 300~G \citep[see, e.g., Figure 3~of][]{asensio_trujillo_hazel08}.
Therefore, for the magnetic fields of up to a few hundreds of Gauss characteristic of spicules and prominences, the Stokes $V$ profiles in the 1083.0~nm line are expected to be mainly antisymmetric.

\begin{figure*}
\centering
\includegraphics[width=0.49\textwidth]{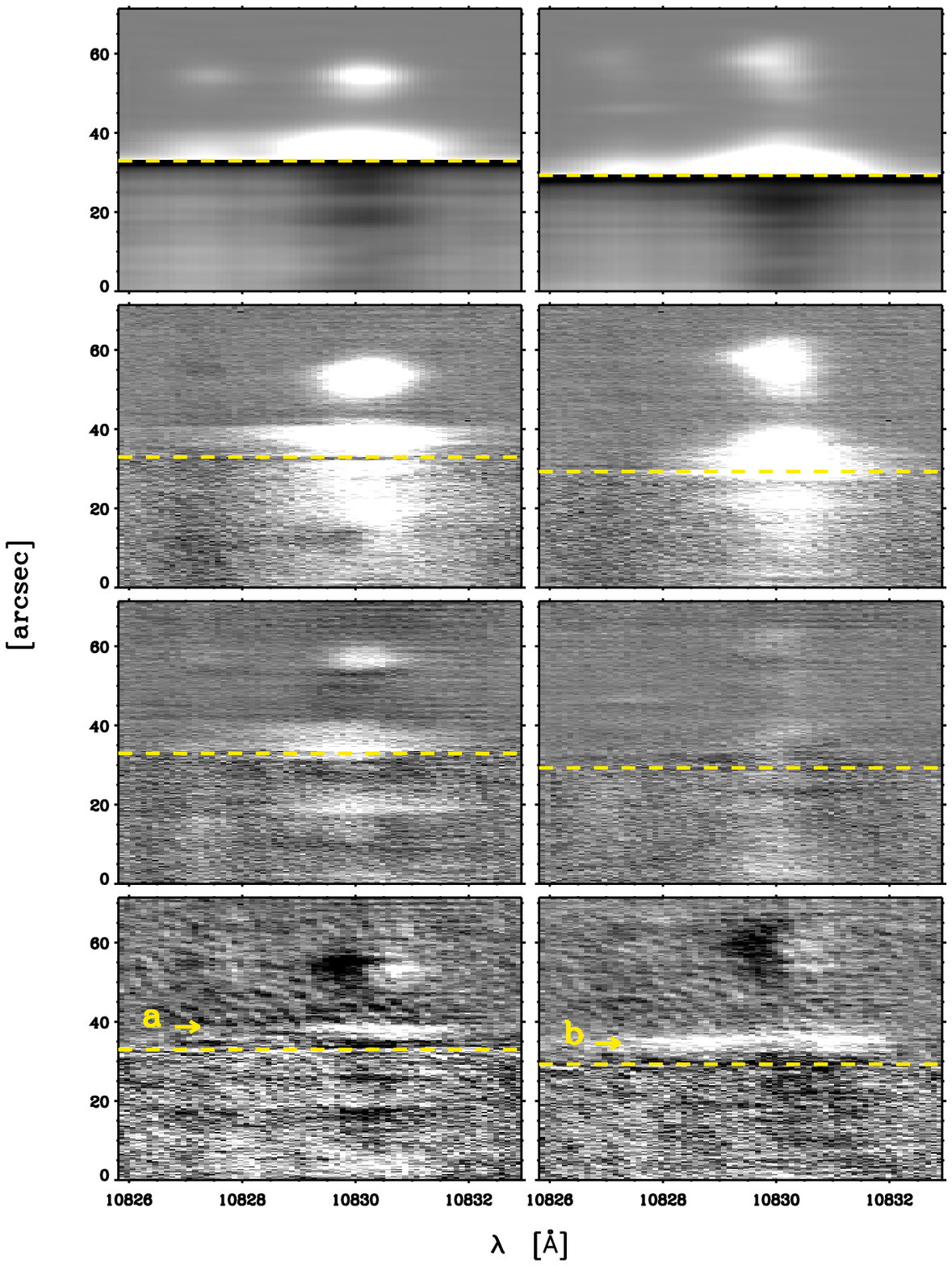}
\includegraphics[width=0.49\textwidth]{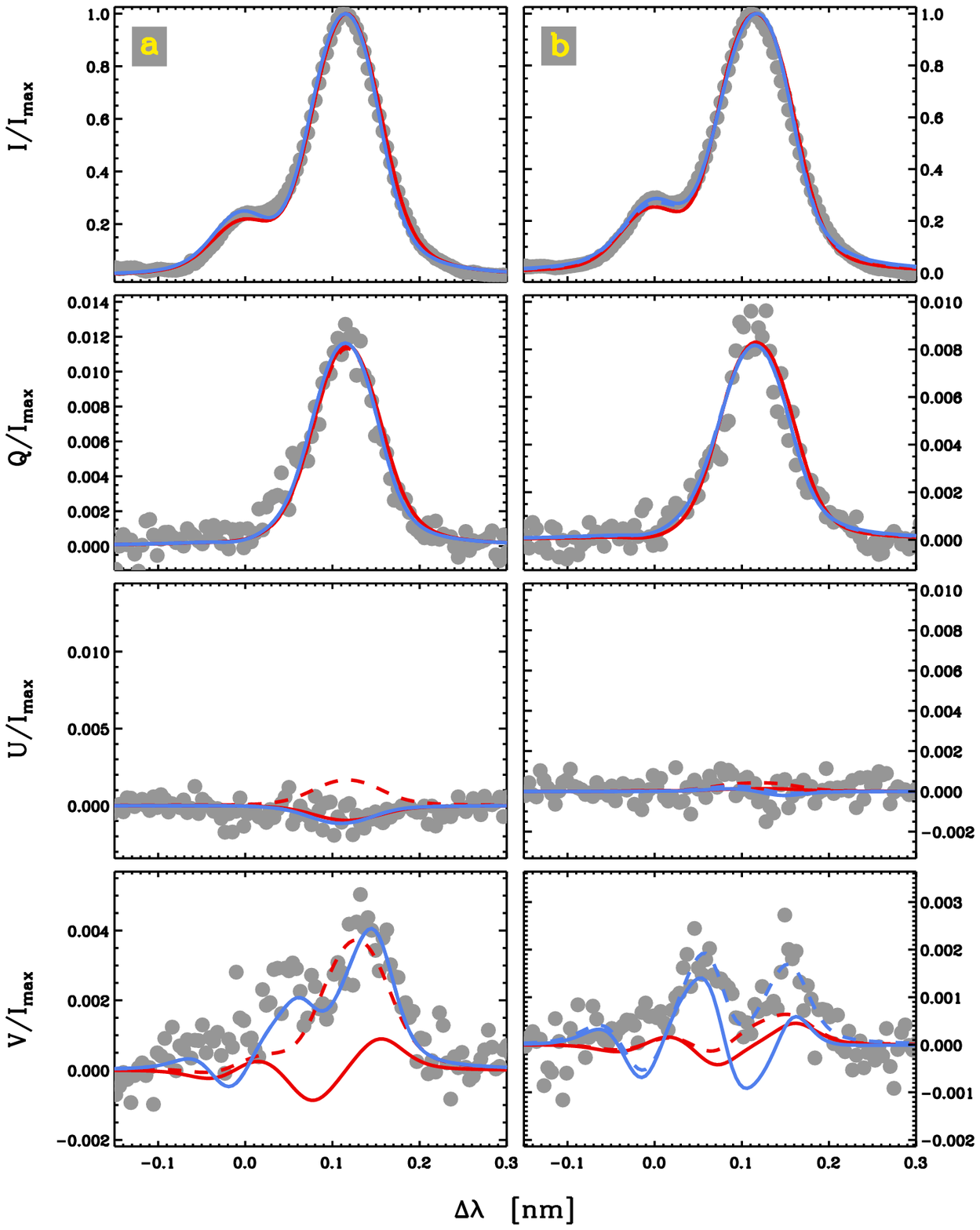}
\caption{Left: Stokes parameters $I$, $Q$, $U$, and $V$ (top to bottom) along the slit at positions 10.8'$''$ (leftmost) and 15.0$''$ (rightmost) in the first and second scan in Figure~3. Dashed yellow lines mark the position of the solar limb (disk, below; off-limb, above). 
The intensity scale on-disk has been highly attenuated for representation purposes. 
Right: line intensity and polarization (dots) at the two positions marked ``a'' and ``b'', and best-fit profiles from inversions with one (red) or two (blue) components along the LOS, in the absence (solid) or presence (dashed) of atomic orientation. The physical parameters of the models are given in Table~1, labeled as $1$ and $2$ for the one-component inversion, in the last case including an ad-hoc orientation of the radiation field, and 3 and 4 for the two-components inversions without and with orientation of the radiation field.
}
\label{fig:mapas_slit1}
\end{figure*}

We report full-Stokes observations of the 1083.0~nm multiplet of \ion{He}{1} in solar spicules and a prominence in a quiet region. 
The observations display anomalous single-lobed Stokes $V$ profiles in the spicular material. 
We show, after a detailed analysis of possible instrumental sources of error, that these profiles are not artifacts, but of solar origin. 
We argue that these profiles are the result of several components along the line of sight (LOS) and that they might also be a signature of atomic orientation.

\section{Observations}
The data analyzed  consist of full Stokes vector spectropolarimetry at the 1083.0~nm spectral region, containing the photospheric \ion{Si}{1} 1082.71~nm line and the chromospheric \ion{He}{1} 1083.0~nm multiplet. 
They were obtained on April 24th, 2011 with the upgraded Tenerife Infrared Polarimeter \citep[TIP-{\sc II};][]{collados_tipII07} instrument attached to the Vacuum Tower Telescope at the Observatorio del Teide. 
We scanned a quiet region at the east limb (90$^\circ$E, 40$^\circ$S), with the slit inclined by about $45^\circ$ with respect to the limb. We simultaneously recorded part of the
solar disk, the spicular material, and a prominence. 
The scanned area was 70$'' \times 42''$ along the slit and in the scan direction, 
respectively. The pixel size was 0\farcs32$\times 0\farcs6$. The time span between
two consecutive slit positions was approximately 30 s and the spectral resolution was 4.5 pm. The 
noise level in polarization was about 10$^{-3}$ in units of the 
maximum intensity, $I_{max}$. The seeing conditions were very good and the adaptive
optics system worked efficiently for the whole observation run. This resulted in a spatial 
resolution  of $\sim 0.6''$ and made it possible that we could scan the same region four times. 

To avoid the appearance of spurious signals we have been extremely careful in the data reduction process. 
From the original raw data, we removed the bias level, applied the flatfield correction, and demodulated the data to obtain the four Stokes parameters using the standard software package available for TIP-II users.
The flatfield correction removed most of the polarized interference fringes, leaving a residual pattern with amplitudes below 10$^{-4}$ $I_{max}$. 

Despite the good seeing conditions and the image stability offered by the adaptive optics system, some occasional image motion is unavoidable within the $4\times 250$~ms of the modulation process which yields spurious polarization \citep[seeing-induced crosstalk, e.g.,][]{lites_87}.
Normally this effect is minimized by the spatio-temporal demodulation scheme \citep{collados99}. 
However, close to the limb observations are extremely sensitive to this effect because of the steep intensity gradients.
As a result, thin stripes of spurious positive and/or negative polarization signal parallel to the limb were apparent in the observations.
%However, the flatfield correction was not enough to remove some features in the polarized continuum.
%We forced the polarized continuum level to zero because we focus on the line polarization.
%We eliminated the seeing-induced crosstalk from Stokes $I$ to the polarized spectrum in a statistical way. 
At this point, the observational procedure with the slit crossing the limb proved to be very helpful. 
The spurious polarization signal along the slit can be modeled as a combination of the first and second derivatives of the intensity, while the actual coefficients of the combination depend on the particular image motion during the observation.
Assuming that the continuum is unpolarized  because linear polarization due to Rayleigh scattering is negligible at such long wavelengths, we fitted the variation of the spurious polarization signals in the continuum along the slit to the first and second spatial derivatives of the intensity, and subtracted the fit.
This correction was applied to all wavelengths using the same coefficients but the intensity derivatives calculated at each wavelength. 
This process was able to remove most of the artifacts but within a very small stripe ($\leq 1''$) around the limb.
As an additional benefit, this procedure further diminished the residual patterns left after flat field correction.

\begin{figure*}
\centering
\includegraphics[width=0.49\textwidth]{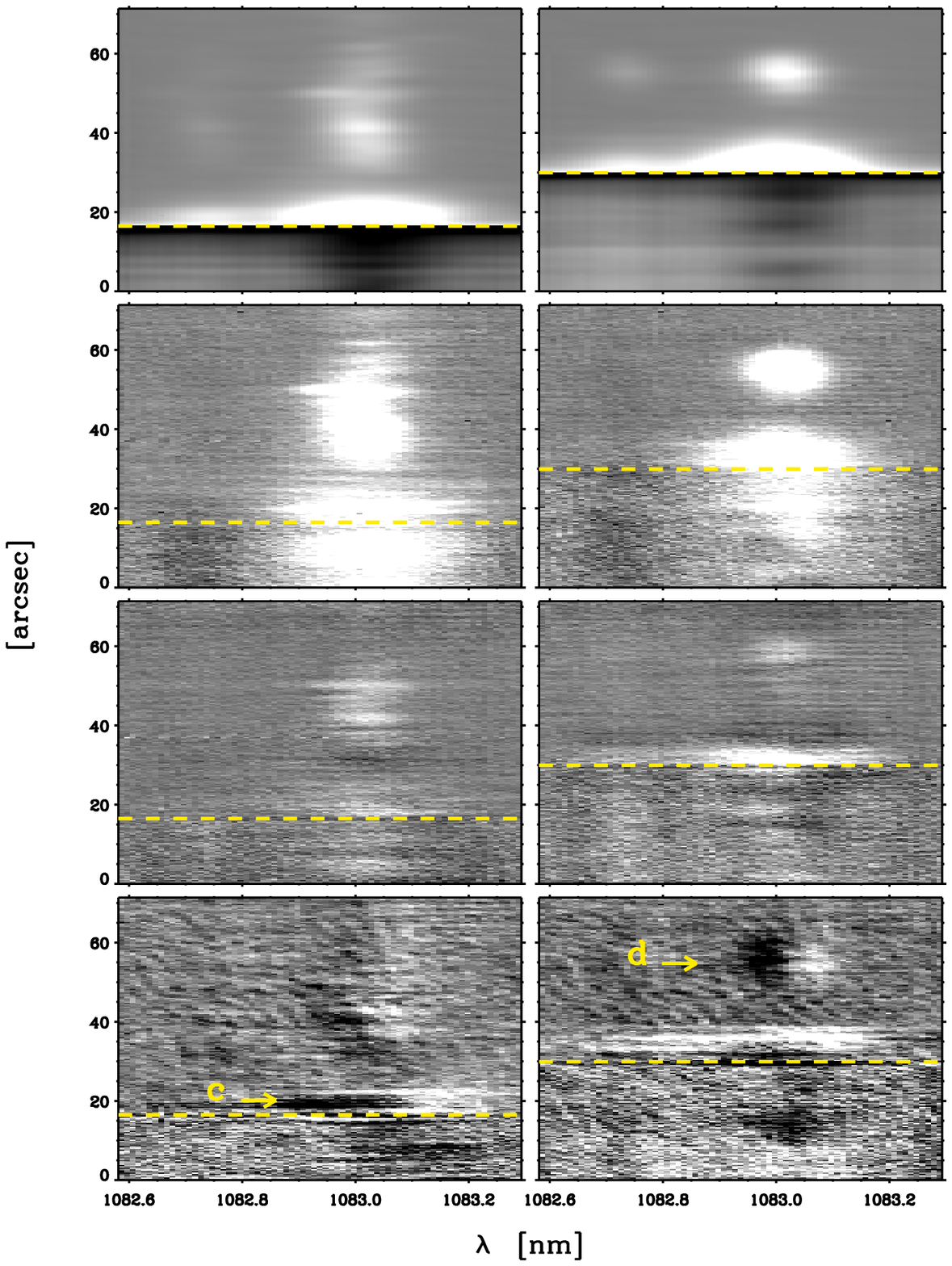}
\includegraphics[width=0.49\textwidth]{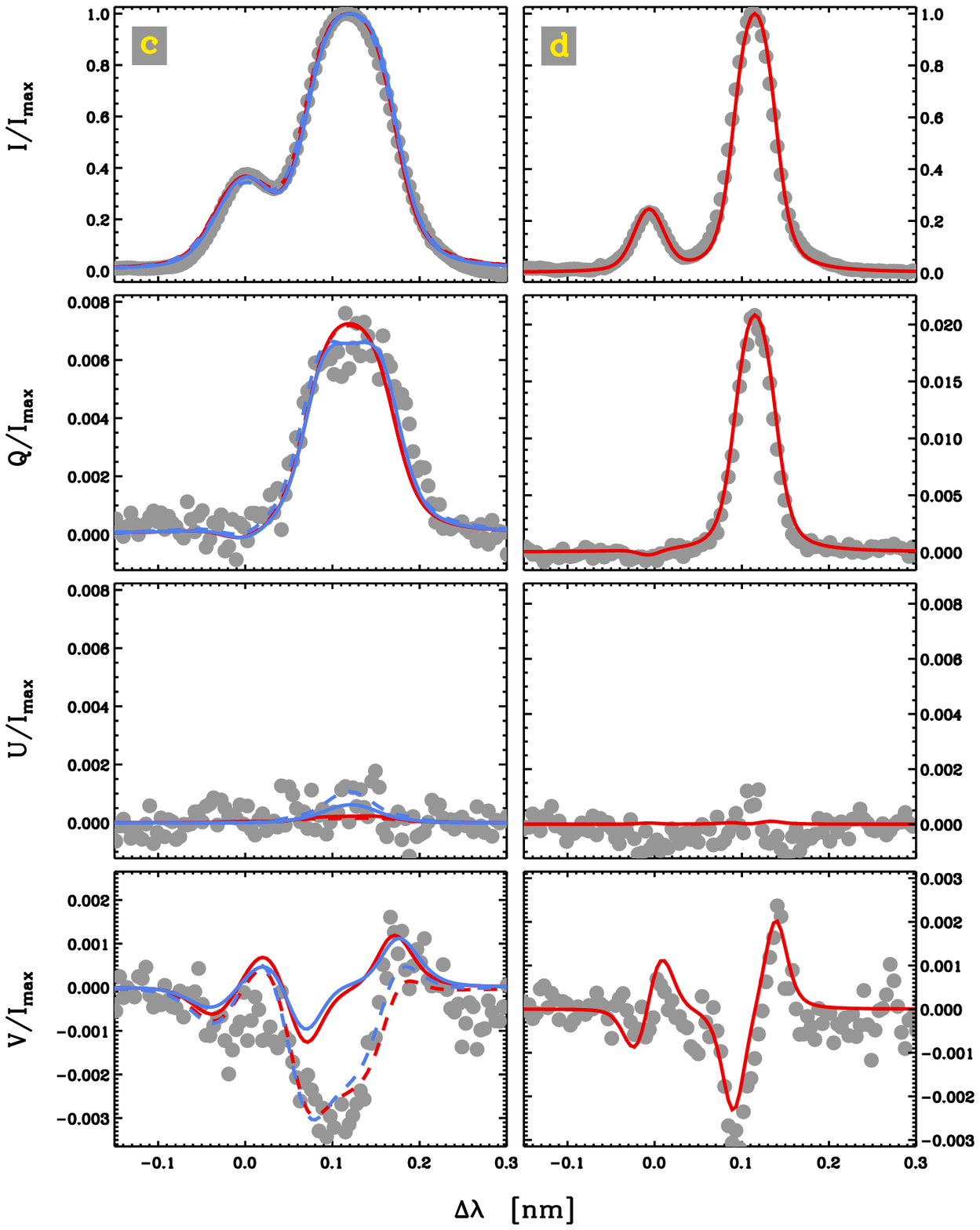}
\caption{Same as Figure \ref{fig:mapas_slit1} but for slit positions 23.4$''$ and 13.2$''$ in the third and first scan in Figure 3.}
\label{fig:mapas_slit2}
\end{figure*}

%We took, from each slit position, the first and second derivatives of the continuum variation along the slit. Then, we 
%correlate these two derivatives with the variation of the polarized continuum along the slit to compite the spurious signal 
%induced by seeing. We then remove the spurious seeing induced polarized spectra. Note that this correction is only feasible if 
%the linb is present in the observations. All observations with the slit parallel to the limb can have spurious signals due to seeing 
%which can not be removed.

Stray light is also an issue when observing across the limb \citep{beck_11}.
At some far off-limb positions with no apparent emission from the prominence a residual absorption spectrum from the integrated disk could be noticed.
From the average of several such pixels we built a stray light spectrum.
The amount of stray light contamination along the slit was then found through correlation with this spectrum, using the telluric line at 1083.2 nm and smoothing the ensuing correlation coefficient maps. 
% The level of stray light thus found increases when approaching the limb, and remains constant (at a negligible level) whithin the disk.
We then subtracted the average off-limb stray light profile times the smoothed correlation coefficient from the Stokes $I$ profiles. This approach proved to be quite effective in removing off-limb stray light. 
%because a) the off-limb telluric line profile was a clear indicator of stray light, and b) the stray light spectrum was slightly shifted in wavelength with respect to the solar limb one.

Figures \ref{fig:mapas_slit1} and \ref{fig:mapas_slit2} display the Stokes profiles of the 1083.0~nm
triplet at four different positions. The intensity profile 
appears in emission above the solar limb, both in the spicules and in the
prominence (at about $25''$ above the limb). 
The \ion{He}{1} multiplet shows characteristic signatures associated to scattering processes
and the Hanle effect inside and outside the disk in Stokes $Q$ and $U$. The
Stokes $V$ signals are very faint on the disk but show clearly antisymmetric
profiles in the prominence (profile d), as expected for this multiplet when the Stokes $V$
signal is dominated by the Zeeman effect. 
However, unexpectedly, we detect single-lobed Stokes $V$ 
profiles off-limb in the spicular material.

\begin{figure}
\centering
\includegraphics[width=\columnwidth,bb=15 30 500 296]{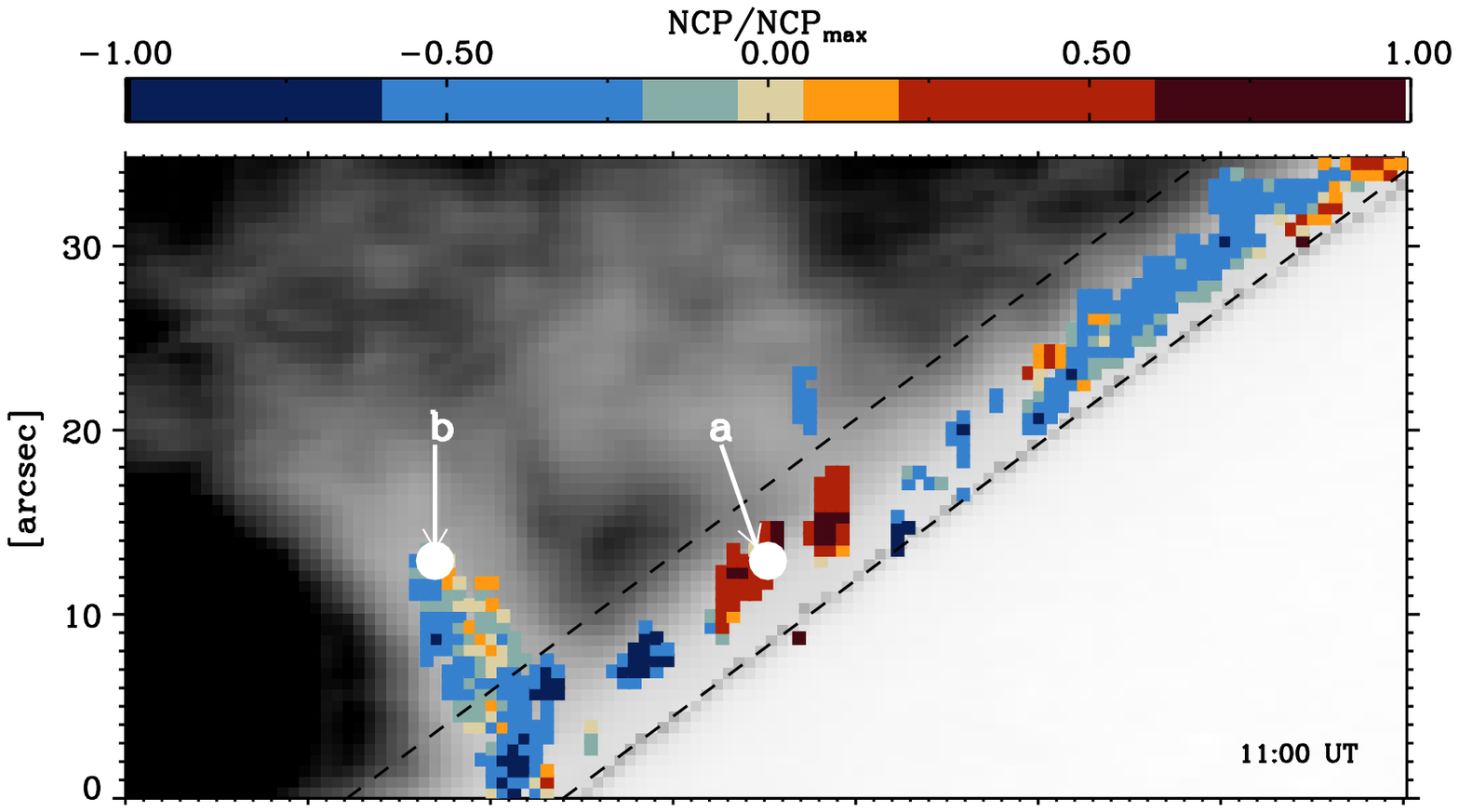}\\
\includegraphics[width=\columnwidth,bb=15 30 500 255]{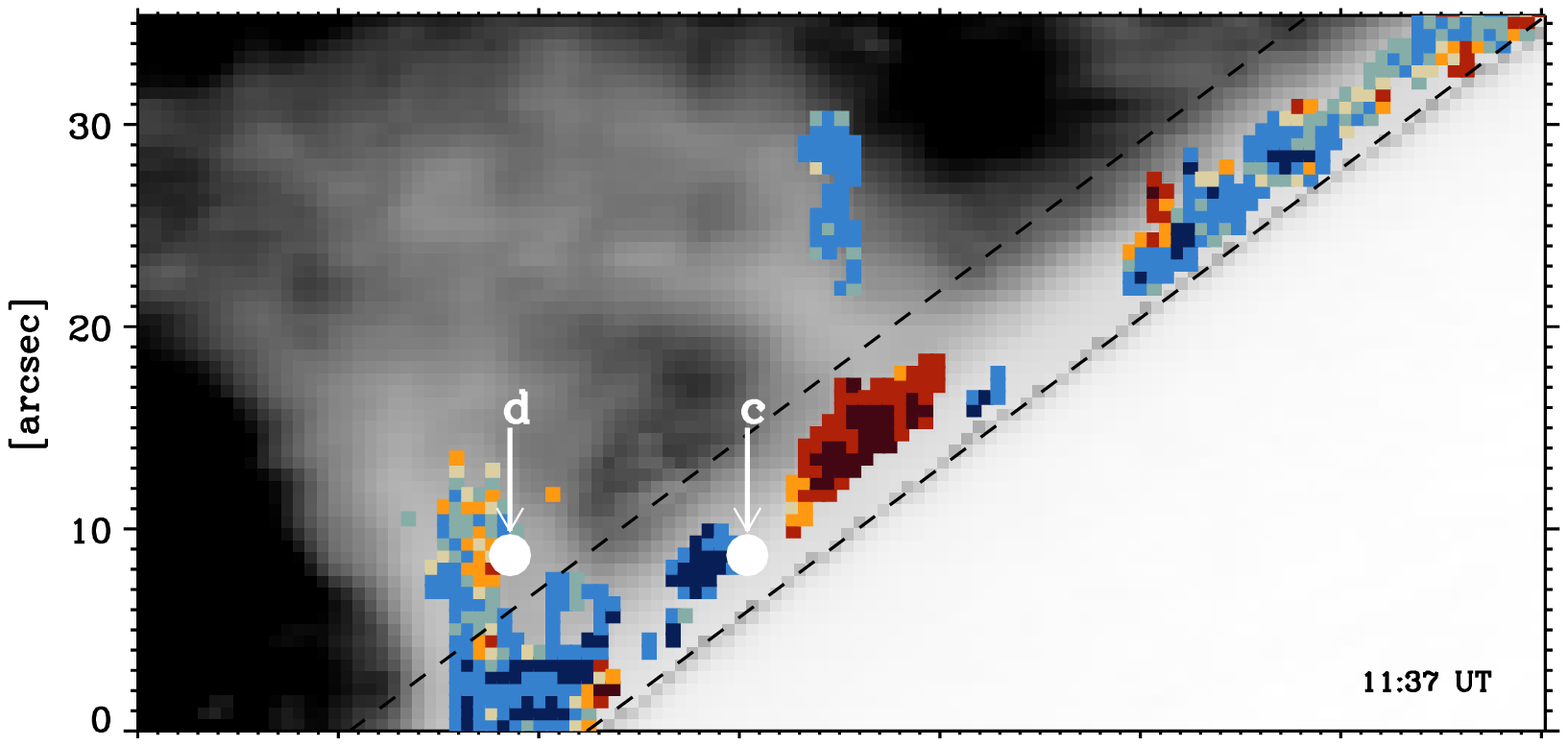}\\
\includegraphics[width=\columnwidth,bb=15 30 500 255]{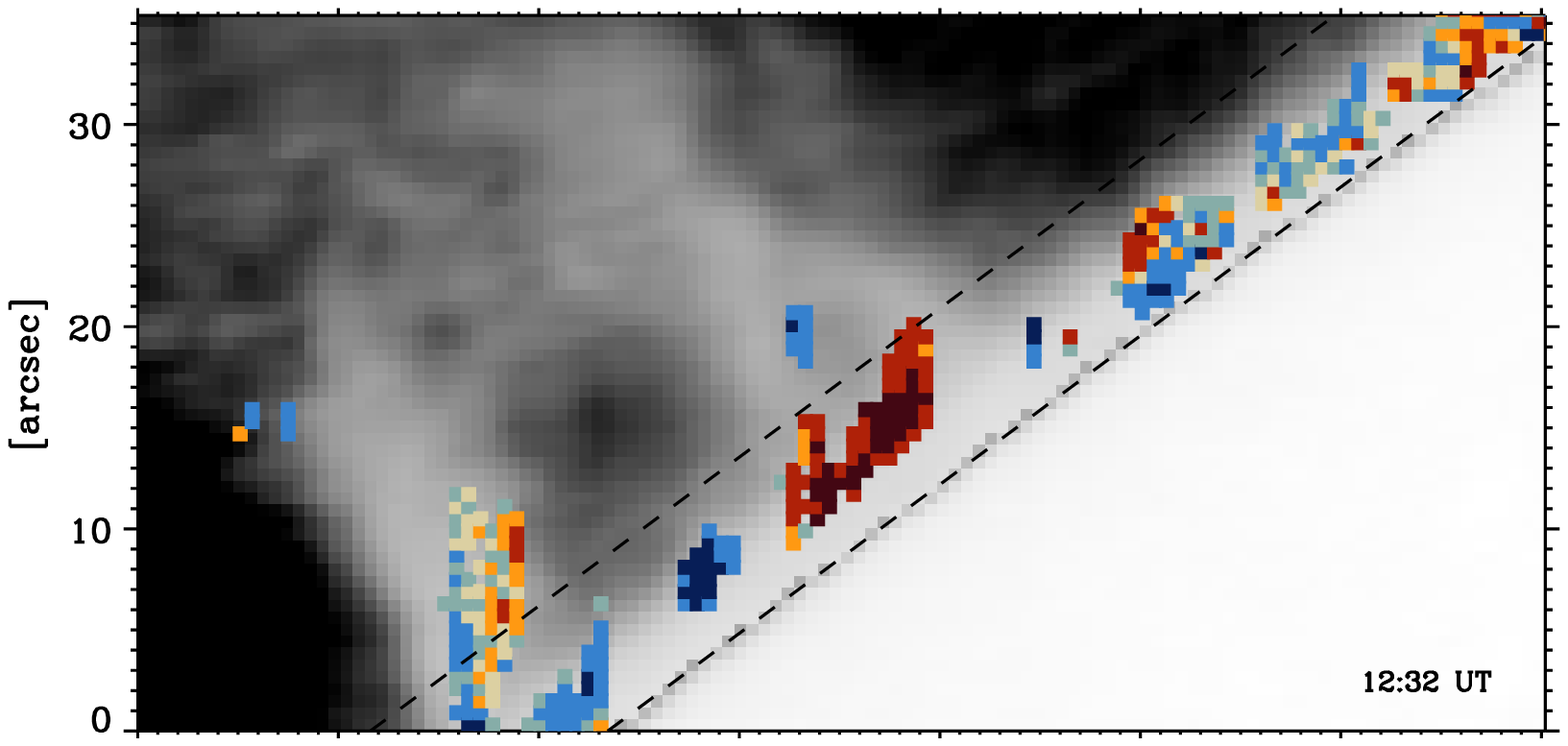}\\
\includegraphics[width=\columnwidth,bb=15 30 500 290]{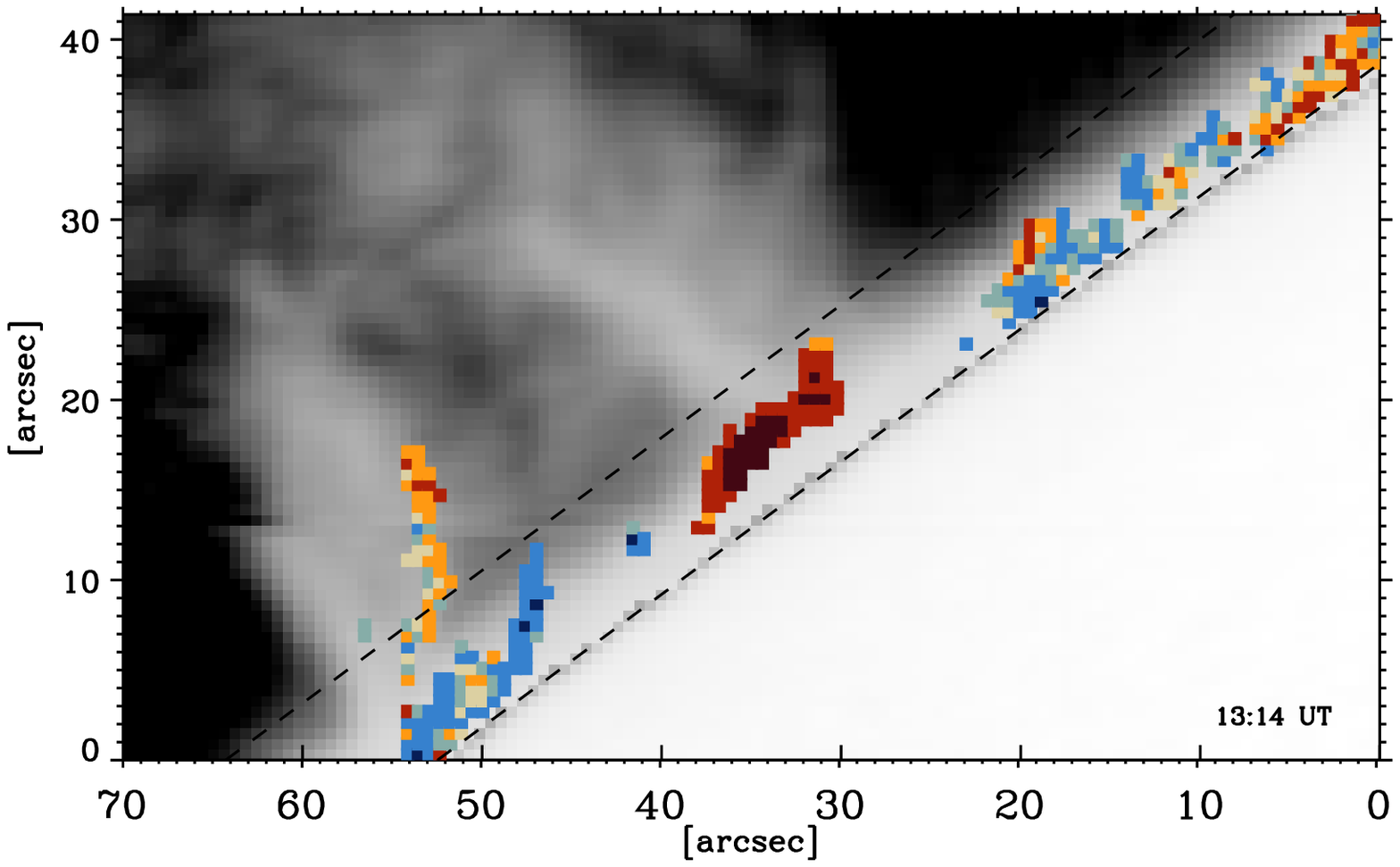}
\vspace{0.3cm}
\caption{Four consecutive scans (horizontal slit; scanning direction from bottom to top) of the observed FOV including spicules and a prominence.
The start time of each scan is displayed at the bottom right corner of the corresponding panel.
The background images show the maximum intensity of the \ion{He}{1} multiplet. The NCP of Stokes $V$ profiles (normalized to its maximum
in the field-of-view) with signals above the noise level is overplotted in a color code.
The dashed lines mark the solar limb and, roughly, the highest altitude of observed spicular material. White circles mark the position of the profiles shown in Figures ~\ref{fig:mapas_slit1} and \ref{fig:mapas_slit2}.}
\label{fig:mapas_int_ncp}
\end{figure}

Firstly, we note that the single-lobed $V$ profiles were already apparent after the standard data reduction process; the thorough analysis described above just served to enhance their presence by eliminating artifacts.
Additional reasons why we think these single-lobed Stokes
$V$ profiles are of solar origin are: 
1) we do not observe any off-limb polarization signal in the telluric line \citep[see Sect. 3.3 in][]{beck_11}, 
2) we also observe Zeeman-like signals at the same slit position, 3) they are located
in patches that are spatially coherent during more than 135 minutes. It seems extremely unlikely that seeing-induced signals appear at the same location in the solar structure during all the observational runs. Figure \ref{fig:mapas_int_ncp} displays the net circular polarization (the Stokes $V$ signal integrated along the 1083.0~nm triplet) 
normalized to its maximum in the field of view (FOV) for pixels with  
circular polarization signals above the noise level. 
Extremely asymmetric Stokes $V$ profiles are quite common in spicules forming patches that are persistent for more than 2 hours of observation.

Regular Stokes $V$ profiles (mostly antisymmetric and similar to profile d) pervade the prominence. They are also found in the spicules, although they have a large degree of asymmetry. Figure \ref{fig:mapas_slit2} shows two examples of antisymmetric profiles. As can be seen, the Stokes $V$ profile belonging to the spicular material is asymmetric (both in area and in amplitude). Moreover, both Stokes $Q$ and $V$ profiles show two peaks and are broader than Stokes $I$, indicating the existence of at least two atmospheric components along the line of sight.

\section{Inversion of the data}

We carried out inversions of the profiles shown in Figures~\ref{fig:mapas_slit1} and \ref{fig:mapas_slit2} with the inversion code HAZEL \citep[HAnle and ZEeman Light, see][]{asensio_trujillo_hazel08} assuming a slab of constant properties, characterized by the optical depth of the red component of the multiplet, $\tau_{\rm red}$, the thermal width, $v_\mathrm{th}$, the macroscopic velocity along the LOS, $v_\parallel$, and the damping constant. Each slab is permeated by a magnetic field with strength, $B$, inclination relative to the vertical, $\theta_B$, and
azimuth, $\chi_B$. The radiation field is assumed to be given by the continuum
illumination at the wavelengths of all transitions considered in the 5-level \ion{He}{1}
atomic model. The center-to-limb variation tabulated by \cite{cox00} is used to compute 
the anisotropy of the radiation field. This simple model let us fit regular profiles as d) in Figure~\ref{fig:mapas_slit2} and recover the physical conditions there (see model d in Table~1), but it is clearly unable to reproduce the $V$ profiles at positions a, b, and c (cf. solid red lines in Figures~\ref{fig:mapas_slit1} and \ref{fig:mapas_slit2}, corresponding to models a$_1$, b$_1$, and c$_1$ in Table~1). In general, the Stokes $V$ profiles in the prominence are similar to profile d) in Figure~\ref{fig:mapas_slit2}, implying that a single magnetic component would be sufficient to fit the observations of a prominence at our spatial resolution and spectropolarimetric accuracy.

Obviously, the Zeeman effect alone is unable to reproduce Stokes $V$ profiles with NCP but several known mechanisms are able to generate it. First, the presence of correlations between velocity and magnetic field gradients along the LOS \citep[][]{illing75} has often been proposed (in photospheric lines) to explain the asymmetries in Stokes $V$ induced by the Zeeman effect 
around magnetic flux concentrations \citep{solanki88,grossman_doerth88,sanchez_almeida89,solanki93}. 
This effect might be of some importance when observing spicules or prominences close to and off the limb, because
the LOS potentially crosses many structures with different velocities and magnetic fields \citep[e.g. the multi-thread prominence models described in][and references therein]{labrosse_10, mackay_10}.
A property of the asymmetric profiles is that the width of the remaining single-lobed Stokes
$V$ profile should be roughly half the width of the Stokes $I$ profile. In our case,
the observed single-lobed Stokes $V$ profile is roughly of the same width as Stokes
$I$, and in some cases even larger. 
% In principle, explaining our observations in terms of gradients turns out
% to be difficult, although we do not discard it completely.

Atomic orientation, i.e., the presence of population imbalances and quantum coherence between the $+M$ and the corresponding $-M$ magnetic sublevels of a given level \citep[][]{fano57}, is in itself a direct source of NCP through imbalanced emission or absorption of $\sigma^\pm$ components. To fit the profiles, we have considered the possible presence of atomic orientation (models a$_{2}$, b$_{2}$, c$_{2}$ in Table 1 and dashed red lines in Figures \ref{fig:mapas_slit1} and \ref{fig:mapas_slit2}). The way to introduce this atomic orientation in the HAZEL code is by an ad-hoc amount of orientation of the incident radiation field (i.e., $J^1_0 / J^0_0$).
We only took into account orientation in this multiplet. This is justified because the 1083.0~nm multiplet can be relatively well approximated by a two-term system. We have nonetheless tested that
the presence of atomic orientation in the other transitions of the model does not
modify our results. With the approximation of a single component and atomic orientation, we obtained an acceptable fit for the position c (although the double peak of Stokes $Q$ is not reproduced), but it was impossible to reproduce the other profiles.

We then performed inversions considering two components along the LOS, each one with different values of the physical parameters and a relative macroscopic velocity between them (models a$_{3}$, b$_{3}$, c$_{3}$ in Table 1 and solid blue lines in Figures \ref{fig:mapas_slit1} and \ref{fig:mapas_slit2}).
Finally, we also allowed for an ad-hoc amount of orientation in the two-component model (models a$_{4}$, b$_{4}$, c$_{4}$ in Table 1 and dashed blue lines in Figures \ref{fig:mapas_slit1} and \ref{fig:mapas_slit2}).
With this modelling we fit fairly well the remaining profiles. The inferred physical parameters are reasonable for these kind of structures: the doppler broadening and the optical thickness have larger values in the spicules than in the prominence, as well as the magnetic field strength (see Table 1). However, the exact magnetic field vector is subject to some known ambiguities whose discussion is beyond the scope of this paper.

In spicules, regular antisymmetric Stokes $V$ profiles are rare; most of them exhibit noticeable NCP and often
show evidence for the presence of (at least) two components. We detect either two peaks in the Stokes $V$ lobes and/or Stokes $Q$ and $U$ that are not compatible with Stokes $V$ with a single configuration of the magnetic field. To fit polarization profiles of solar spicules one therefore needs more complex models than that of a single component. Probably, one should think on a modeling that takes into account the transfer of polarized radiation and a source not only of atomic alignment (hence $J^2_0/J^0_0$) but of atomic orientation (hence $J^1_0/J^0_0$).

\begin{table}
\caption{Inferred physical parameters of the inversion models}
\centering
\begin{tabular}{clrrrrcc}
\hline\hline
%model & $\tau_{\rm red}$ & $B$ (G) & $\theta_B(^\circ)$ & $\chi_B(^\circ)$ & $v_{\rm th}$ (km~s$^{-1}$) & $v$ & . \\
model & $\tau_{\rm red}$ & $B$ & $\theta_B$ & $\chi_B$ & $v_{\rm th}$ & $v_\parallel$ & $v$ \\
 & & (G) & $(^\circ)$ & $(^\circ)$ & \multicolumn{2}{c}{(km s$^{-1}$)} & \\
\hline
a$_1$ & 0.83 & 18  &  93 &  58 & 11.0  & -1.5  &      \\
a$_2$ & 0.83 & 11  & 87  & 57  & 11.0  & -1.5  & 0.05     \\
a$_3$ & 0.78  & 79  & 38  & 4  & 9.0  & -0.8  &      \\
          & 0.54 & 179 &  94 &  109  & 10.4  & -2.7  &       \\
a$_4$ & 0.78  & 57  & 36  & 3  & 9.0  &  -0.5 &   0.013   \\
          &  0.54 &  93 & 93  &  119  & 10.4  &  -4.4  &   0.013    \\
 \hline                                   
b$_1$ & 1.5  & 7  & 92  & 46  & 11.5  & -1.5  &      \\
b$_2$ & 1.5  & 5  & 92  & 46  & 11.5  & -1.5  & 0.01     \\
b$_3$ & 1.0  & 38  & 41  & 7  & 8.4  &  0.3 &      \\
          &  0.7 &  60 &  88 &  131 &  10.3 &  -4.7 &       \\
b$_4$ & 0.9  & 35  & 41  & 6  & 9.2  & 0.3  &    0.006  \\
          &  0.6 & 71  &  88 &  130 &  11.0 &  -4.8 &   0.006   \\
\hline                                   
c$_1$ & 3.0  & 19  & 40  & 3  & 10.5  & -0.6  &      \\
c$_2$ &  3.0 & 21  & 40  & 3  & 10.9  & -0.6  & -0.004     \\
c$_3$ & 1.7  & 29  & 93  & 85  & 10.0  & -0.1  &      \\
          &  1.3 &  23  &  86 &  31 &  12.3 &   &  -0.1     \\
c$_4$ & 1.4  & 75  & 14  & 25  & 10.1  & -0.7  & -0.006     \\
          &  1.2  &  22 &  65 &  30 &  13.0 &  -0.6 &  -0.006    \\
\hline                                   
d        &  1.6 & 28  & 44  & 3  & 5.5  & -1.8  &      \\
\hline \hline
\end{tabular}
\end{table}

\section{Generation of atomic orientation}

As stated in Sect.\,1, the A-O mechanism can generate a very small amount of atomic orientation that is already taken into account in the modeling by HAZEL. Likewise, the generation of a sufficient amount of atomic orientation through the presence of electric fields \citep{favati87} as suggested by \cite{lopezariste_halpha05} \citep[see also][]{casini_rafa_06, casini_rafa_08} for the H$\alpha$ line in solar prominences can also be discarded for this multiplet since it is not hydrogenic.

Atomic orientation may be induced by differential excitation of the $\sigma^+$ and $\sigma^-$ components of the transition. This can be achieved by a) illuminating the atomic system with circularly polarized radiation or b) by spectrally splitting the transition components (e.g., in the presence of a magnetic field) and differentially illuminating the $\sigma^\pm$ components. 

In the case a), considering a two-level atom with an unpolarized lower level, and assuming complete frequency redistribution (CRD), the atomic orientation generated in the upper level (hence, the NCP of the scattered radiation) is proportional to (Landi Degl'Innocenti \& Landolfi 2004)
\begin{equation}\label{eq_j10}
v= \frac{\bar{J}^1_0}{\bar{J}^0_0}= 
\sqrt{\frac{3}{2}} \frac{\int_0^\infty\int_{0}^1 V(\nu,\mu)\mu d\mu \,\phi(\nu) d\nu}{\int_0^\infty\int_{0}^1 I(\nu,\mu) d\mu \; \phi(\nu) d\nu},
\end{equation} 
where $\phi(\nu)$ is the line absorption profile. 

\begin{figure}
\centering
\includegraphics[width=\columnwidth]{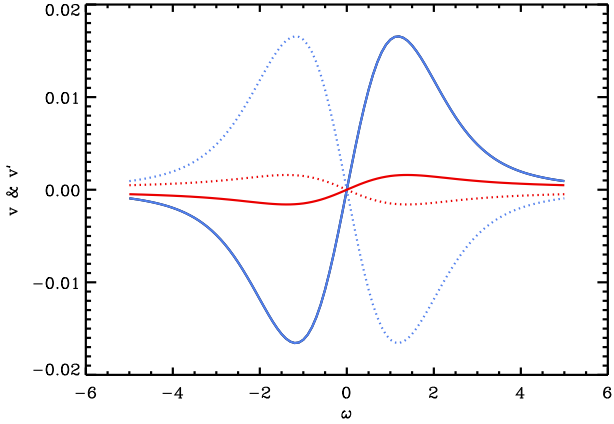}
\caption{Variation of $v$ (blue lines) and $v'$ (red lines) with the adimensional velocity $\omega=v_z \lambda_0/(wc)$ ($v_z$ is the vertical velocity of the upward moving scatterers, $c$ the speed of light, $w$ the width of the incident spectral line).
The incident photospheric radiation field is assumed to be $I=I^0(1-\mu+u\mu)[1-d \exp(-(\lambda-\lambda_0)^2/w^2)]$. 
We have considered an absorption ($a=0.3$), limb darkened ($u=0.51$) line, at $\lambda_0$=1083.0~nm, with a width $w=28.9$~pm (corresponding to 8 km~s$^{-1}$).
The parameter $v$ is calculated assuming that the underlying photospheric field is permeated by a uniform 1~kG field; $v'$ is calculated assuming that the scatterers are in the presence of a 100~G field.
The width of the scatterers absorption profiles $\phi$ has been taken to be similar to the incident line.
The solid and dotted lines correspond to opposite magnetic field polarities.
Positive $\omega$ for upward velocities.
As a reference, $\omega=1$ corresponds to $v_z=8$~km~s$^{-1}$.}
\label{fig:j10}
\end{figure}

As a possible scenario for this, consider an ensemble of scatterers in a spicule or a prominence well above the solar surface illuminated from an underlying magnetized photosphere. If the incident Stokes-$V$ profile is antisymmetric and there is no Doppler shift between the radiation field and the atomic system, $v\equiv 0$ and no atomic orientation can be induced.
However, if the scatterers have a collective motion with respect to the surface, the line profiles $\phi$ in Equation~(\ref{eq_j10}) are Doppler shifted with respect to the incoming radiation field ($I$ and $V$) and $v$ does not cancel out in general. Clearly, if the incident radiation has NCP, that contribution adds to $v$.
%atomic polarization is generated even if there is no Doppler shift between the atom and the incident radiation.
%This effect will be maximum for a Doppler shift of the order of half the separation between the $V$ peaks.
Figure~\ref{fig:j10} shows the amount of $v$ generated as a function of the velocity, assuming upwards motion, for an illumination with a Gaussian absorption line, using values characteristic of the He~{\sc i} 1083.0~nm line in an active region. The Stokes $V$ incident light has no NCP.

In case b), the uneven pumping of the $\sigma$ components of a two-level atom (unpolarized lower level, CRD), by a spectrally structured radiation field is characterized by the ratio 
\begin{equation}\label{eq_phi}
v'=\frac{\int_0^\infty \int^1_0 I(\nu) d\mu \Phi^{10}_0(\nu) d\nu}{\int_0^\infty \int^1_0 I(\nu)  d\mu \Phi^{00}_0(\nu)  d\nu}.
\end{equation}
The frequency integrals are performed over the generalized profiles \citep{landi_landolfi04}
\begin{multline}\label{eq_phi2}
\Phi^{00}_0(\nu)=\sum_{M_uM_\ell}\frac{1}{3}S_q^{J_\ell J_u}(M_\ell M_u) \phi_{M_u M_\ell}(\nu), \\
\Phi^{10}_0(\nu)=\sum_{M_u M_\ell}\frac{M_u}{\sqrt{3J_u(J_u+1)}} S_q^{J_\ell J_u}(M_\ell M_u) \phi_{M_u M_\ell}(\nu),
\end{multline}
where $S_q^{J_\ell J_u}(M_\ell M_u)$ with $q=M_\ell-M_u$ ($q=0, \pm 1$) is the relative strength of the transition between the sublevels $M_u$ and $M_\ell$, and the profile components $\phi_{M_u M_\ell}$ are centered at the frequency $\nu_{M_u M_\ell}=\nu_0+\nu_L(g_u M_u-g_\ell M_\ell)$ ($\nu_L$ is the Larmor frequency and $g_{u, \ell}$ the Land\'e factors of the upper and lower levels, respectively).
%In a two-level atom with an unpolarized lower level, and assuming CRD, the atomic orientation generated in the upper level is proportional to $v'$. 

As a scenario for this case consider that the underlying photosphere is quiet, but that the scatterers are moving (upwards) in the presence of a magnetic field.
The denominator of Equation (\ref{eq_phi}) is the mean intensity integrated over the line profile taking into account the magnetic splitting of the line components.
The generalized profile $\Phi^{10}_0$ is odd with respect to the central frequency of the transition $\nu_0$ \citep[more generally, $\Phi^{KK'}_0(\nu_0+\Delta)=(-1)^{K+K'}\Phi^{KK'}_0(\nu_0-\Delta)$; see][]{landi_landolfi04}.
Therefore, if the intensity profile of the incident radiation is symmetric around the central frequency $\nu_0$, the integral in the numerator of equation (\ref{eq_phi}) vanishes. If the profiles are Doppler shifted with respect to the incoming intensity, $v'$ varies as in figure \ref{fig:j10} (red curve). 

Dynamic generation of atomic orientation has been considered to qualitatively explain asymmetric $V$ profiles in chromospheric lines \citep{trujillo+93}, but its presence in active regions was discarded by \cite{socas_trujillo_ruiz00b}.
Yet, the physical and dynamical conditions to be expected in spicules and prominences could be more favorable for the generation of atomic orientation by the processes just described.
First, velocities of the order of 8-10~km~s$^{-1}$ are easily found in our data from Doppler shifts and from H$\alpha$ movies.
Both scenarios require that the He~{\sc i} 1083.0~nm line is formed in the underlying atmosphere.
% The formation of this line is not completely understood \citep[][]{centeno+08}, but it is known that 
% it is extremely weak or absent in most (quiet) areas of the solar disk; 
% it appears in moderately magnetized regions of the Sun. 
In our observations the 1083.0~nm line is in absorption within the solar disk in Figures~\ref{fig:mapas_slit1} and \ref{fig:mapas_slit2}.
The large values of $v$ that may be achieved in case a) scenario (see Figure~\ref{fig:j10}) require illumination from underlying relatively strong ($\sim$1~kG), unipolar regions. 
However, prominences and filaments form along the neutral line of bipolar active regions, which would yield to cancelations and hence, far smaller values of $v$ and atomic orientation.
More likely is scenario b), in which spicules and low prominence regions are magnetized at the level of $\sim$100~G, values easily found from the inversions.

\section{Conclusions}

We have shown that Stokes $V$ profiles of He\,{\sc i} 1083 nm can have a significant amount of NCP in spicular fields. Surprisingly, this type of signals have not been found in previous spectropolarimetric studies of spicules using this multiplet \citep{socas_elmore05, trujillo_merenda05, centeno+10}.
This may be due to a particularity of the region observed here, or else it could have been that the large values of the NCP might have been mistaken for observational artifacs. To avoid this possible source of confusion, we found our observational strategy of observing with the slit across the limb very advantageous, because it allows a correction for several artifacs caused by image motion or stray light.

These profiles with extreme NCP (mostly single lobed) are explained by the presence of two magnetic components along the LOS (although we expect the reality to be more complex, with a myriad of magnetic and velocity fields crossing the LOS) and perhaps, some amount of atomic orientation ($\sim$ 0.6\%-1\%) generated by dynamic proceses in the presence of magnetic fields.

\begin{acknowledgements}
We thank M. Collados and 
% for very helpful discussions during the data reduction process and analysis, and
J. Trujillo Bueno for useful discussions. 
Based on observations in the VTT operated on the island of Tenerife by the KIS in the Spanish Observatorio del Teide of the Instituto de Astrof\' isica de Canarias.  
Research supported by the Spanish MEC 
under the grants AYA2010-18029 and Consolider-Ingenio 2010 CSD2009-00038.
\end{acknowledgements}

% \bibliographystyle{apj}
% \bibliography{ms.bib}

\end{document}